%% file: Gluino_final_4.tex
\documentclass[aps,prl,twocolumn,superscriptaddress,showpacs]{revtex4}

\usepackage{graphicx}
\usepackage{dcolumn}
\usepackage{bm}

\begin{document}

\newlength{\ogltemp}

\def\met{\settowidth{\ogltemp}{\big\slash}%
\hspace{0.5mm}\big\slash\hspace{-1.4\ogltemp}{E_T}%
}
\def\gev{\,\,\mathrm{GeV}\,}
\def\tev{\,\,\mathrm{TeV}\,}
\def\mev{\,\,\mathrm{MeV}\,}
\def\gevc{\,\,\mathrm{GeV}/c\,}
\def\gevm{\,\,\mathrm{GeV}/c^2\,}


\title{A Search for Scalar Bottom Quarks from Gluino Decays in $\bar{p}p$ Collisions at $\sqrt{s}=1.96$~TeV}

%


\input{July_2005_Authors_and_Siegrist.tex}

\vfill\eject

\date{\today}


\begin{abstract}
We searched for scalar bottom quarks in $156$~pb$^{-1}$ of
{\bf $\bar{p}p$} collisions at {\bf $\sqrt{s}=1.96$}~TeV recorded
by the CDF II experiment at the Tevatron. Scalar bottom quarks
can be produced from gluino decays in R-parity conserving models
of supersymmetry when the mass of the gluino exceeds that of the
scalar bottom quark. Then, a scalar bottom quark can decay into a
bottom quark and a neutralino. To search for this scenario, we
investigated events with large missing transverse energy and at
least three jets, two or more of which were identified as
containing a secondary vertex from the hadronization of $b$
quarks. We found four candidate events, where $2.6\pm 0.7$ are
expected from standard model processes, and placed $95\%$
confidence level lower limits on gluino and scalar bottom quark
masses of up to $280$ and $240\gevm$, respectively.

\end{abstract}

\maketitle


Despite its extraordinary success, the standard model of
elementary particles and interactions (SM) is incomplete since it
does not explain the origin of electroweak symmetry breaking and
the gauge hierarchy problem~\protect\cite{Gildener:1976}. One
proposed extension of the SM, supersymmetry
(SUSY)~\protect\cite{Martin:1997ns}, solves these problems.
R-parity~\protect\cite{Martin:1997ns} conserving SUSY models also
provide a prime candidate for the dark matter in the
cosmos~\protect\cite{Munoz:2003gx}, the stable lightest
supersymmetric particle (LSP). SUSY is a space-time symmetry that
relates particles of different spin by introducing a boson SUSY
partner for each SM fermion and vice versa. For example, the
left-handed and right-handed quarks have scalar partners denoted
$\tilde{q}_L$ and $\tilde{q}_R$ which can mix to form scalar
quarks (squarks) with mass eigenstates $\tilde{q}_{1,2}$. Several
models~\protect\cite{Bartl:1994bu} predict that this mixing can be
substantial for the scalar bottom (sbottom), yielding a sbottom
mass eigenstate ($\tilde{b}_1$) significantly lighter than other
squarks. At the Tevatron center-of-mass energy of
$\sqrt{s}=1.96\tev$, the predicted gluino ($\tilde{g}$, the spin-1/2 partner
of the gluon) production cross section is almost an order of
magnitude larger than that of a sbottom of the same
mass~\protect\cite{Beenakker:1996ch}. Therefore, if sufficiently
light, sbottom quarks could be copiously produced through the
decay of the gluino into a sbottom and a bottom quark since the
gluino preferably decays into a squark-quark
pair~\protect\cite{Martin:1997ns}.
A sbottom in the mass range accessible at the
Tevatron is expected to decay predominantly into a bottom quark
and a lightest neutralino which is often assumed to be the LSP.

Previous searches for direct
sbottom~\protect\cite{Affolder:1999wp} or gluino
production~\protect\cite{Affolder:2001tc} at the Tevatron
placed upper limits on the masses of these particles. In this
letter, we describe the first search for $\tilde{b}_1$ from
$\tilde{g}$ decays in $156$~pb$^{-1}$ of data collected between
July 2002 and September 2003 by the Collider Detector at Fermilab
(CDF II) at the Tevatron. We assume R-parity conservation and,
therefore, that $\tilde{g}$ are produced in pairs. We consider a scenario
where the branching fractions $\tilde{g} \rightarrow b\tilde{b}_1$
and $\tilde{b}_1 \rightarrow b\tilde{\chi}_1^0$ are $100\%$. We
assume a $\tilde{\chi}_1^0$ mass of $60\gevm$ which is above the
current limits from LEP~\protect\cite{neutralino_mass}.
The signature of our search is a final state with four $b$-jets from
the hadronization of the $b$-quarks, an imbalance in energy in the
transverse plane to the beam (``missing transverse energy" or
$\met$) from the two LSPs escaping detection, and no isolated
leptons with large transverse momentum. We expect that these
events could be observed in inclusive three jet events with large
$\met$. After describing the experiment, we will introduce our
event selection, major backgrounds, systematic uncertainties and
results.

CDF II is a multipurpose detector described in detail
elsewhere~\protect\cite{CDFII}. We use a cylindrical coordinate
system around the beam axis in which $\theta$ and $\phi$ are the
polar and azimuthal angles respectively.
The pseudo-rapidity~$\eta$ is defined as $\eta =
-\ln(\tan(\frac{\theta}{2}))$. The transverse momentum and energy
are defined as $p_T=p \sin\theta$ and \mbox{$E_T=E \sin\theta$},
respectively. The tracking system consists of a cylindrical
open-cell drift chamber and silicon microstrip detectors in a
$1.4$~T magnetic field parallel to the beam axis.
The silicon detectors provide precise tracking information for
$|\eta|<2$ and are used to detect displaced
vertices~\protect\cite{SVXII}. The drift
chamber~\protect\cite{COT} surrounds the silicon detectors and has
maximum efficiency for $|\eta|<1$. The energies of electrons and
jets are measured in calorimeters which cover $|\eta|<3.6$. Muons
are identified by drift chambers located outside the calorimeter
volume which extend to $|\eta|<1.5$.

Jets are reconstructed from the energy depositions in the
calorimeter cells using an iterative cone jet clustering
algorithm~\protect\cite{Arnison:1983rn}, with a cone size of
radius \mbox{$R = \sqrt{(\Delta\phi)^2+(\Delta \eta)^2} = 0.4$}.
Energy corrections are applied to account for effects that can
distort the measured jet energy such as non-linear calorimeter
response, underlying events, or the position of the primary
vertex.
The $\met$ is defined by,
$\met = -\sum_{i} E_T^i{\bf \hat{n}_i}$,
where ${\bf \hat{n}_i}$ is a unit vector perpendicular to the beam axis and pointing at the
i$^{th}$ calorimeter tower, and $E_T^i$ the transverse energy therein.
Candidate electrons must have a track associated with a cluster in the
electromagnetic calorimeter with $E_T>10\gev$, an electromagnetic
to hadronic energy ratio, and shower shape compatible with an
electron~\protect\cite{Rott:2004nz}. Candidate muons are
identified with $p_T>10\gevc$ tracks reconstructed in the tracking
system which extrapolate to a track segment in the muon chambers,
and leave an energy consistent with a minimum ionizing particle in
the calorimeter~\protect\cite{Rott:2004nz}. To increase the
lepton efficiency (including $\tau$), candidate leptons are also
identified as isolated tracks with $p_T>10\gevc$.

\begin{figure*}[t]
\includegraphics[width=5cm]{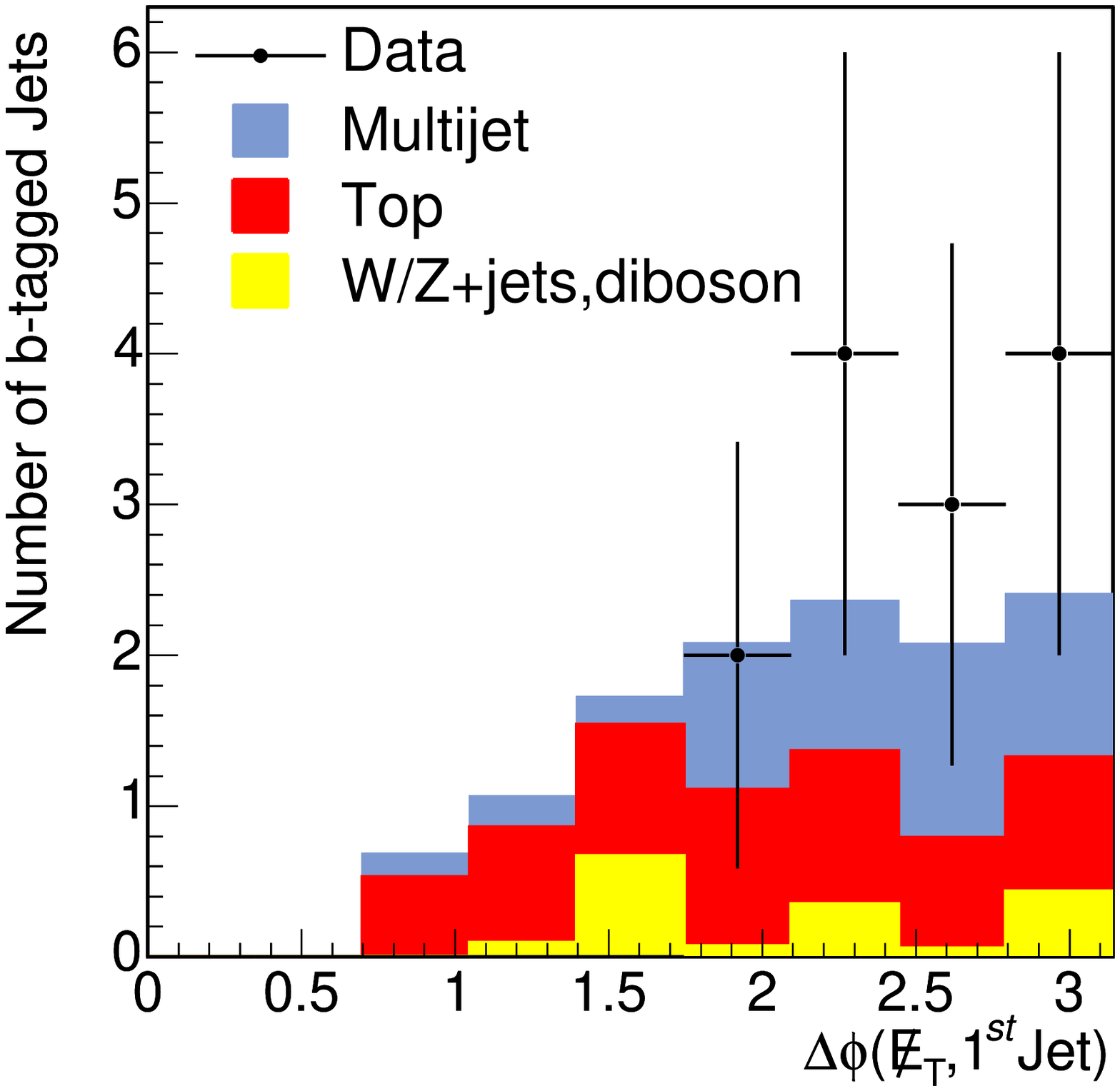}
\includegraphics[width=5cm]{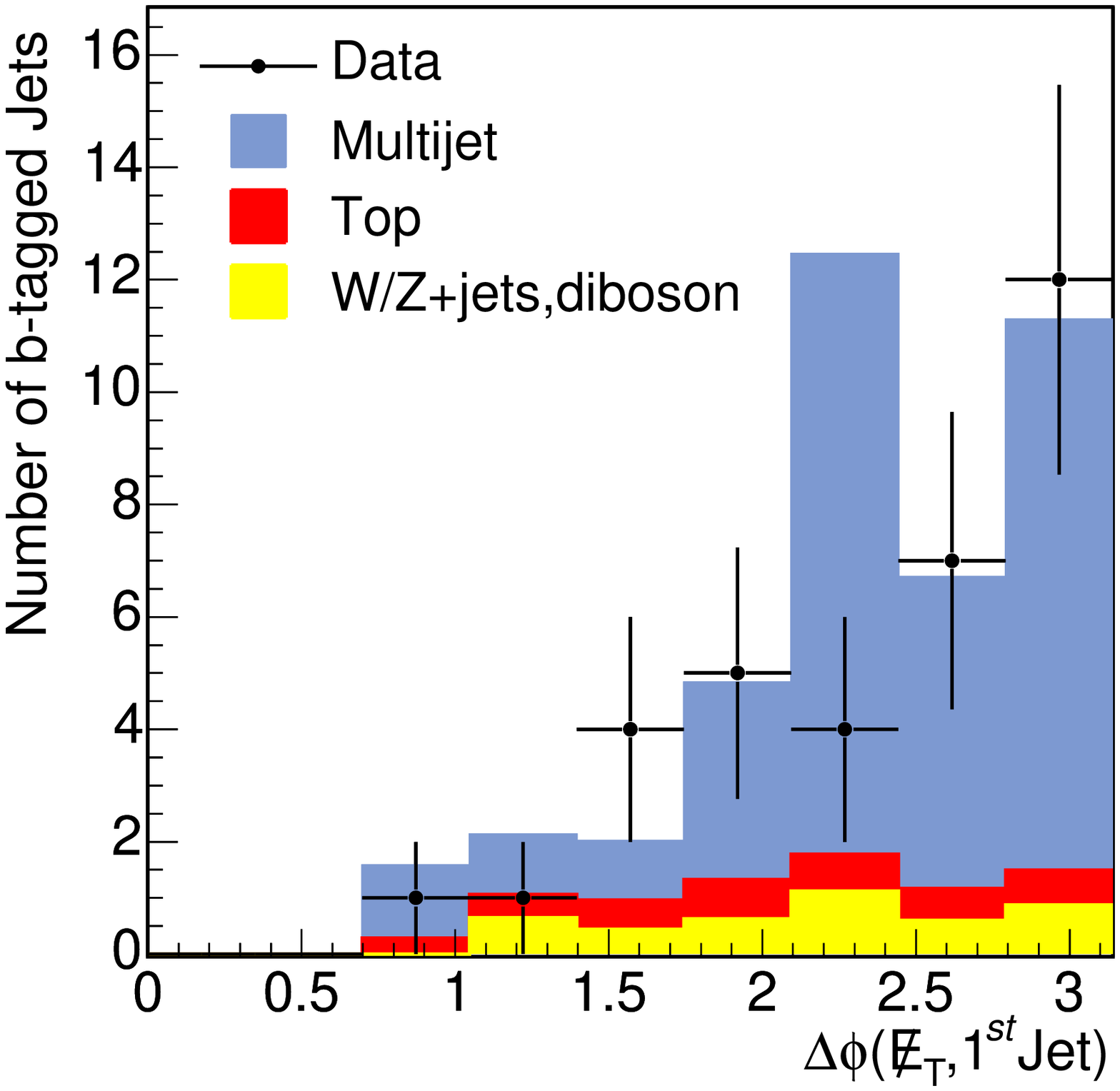}
\includegraphics[width=5cm]{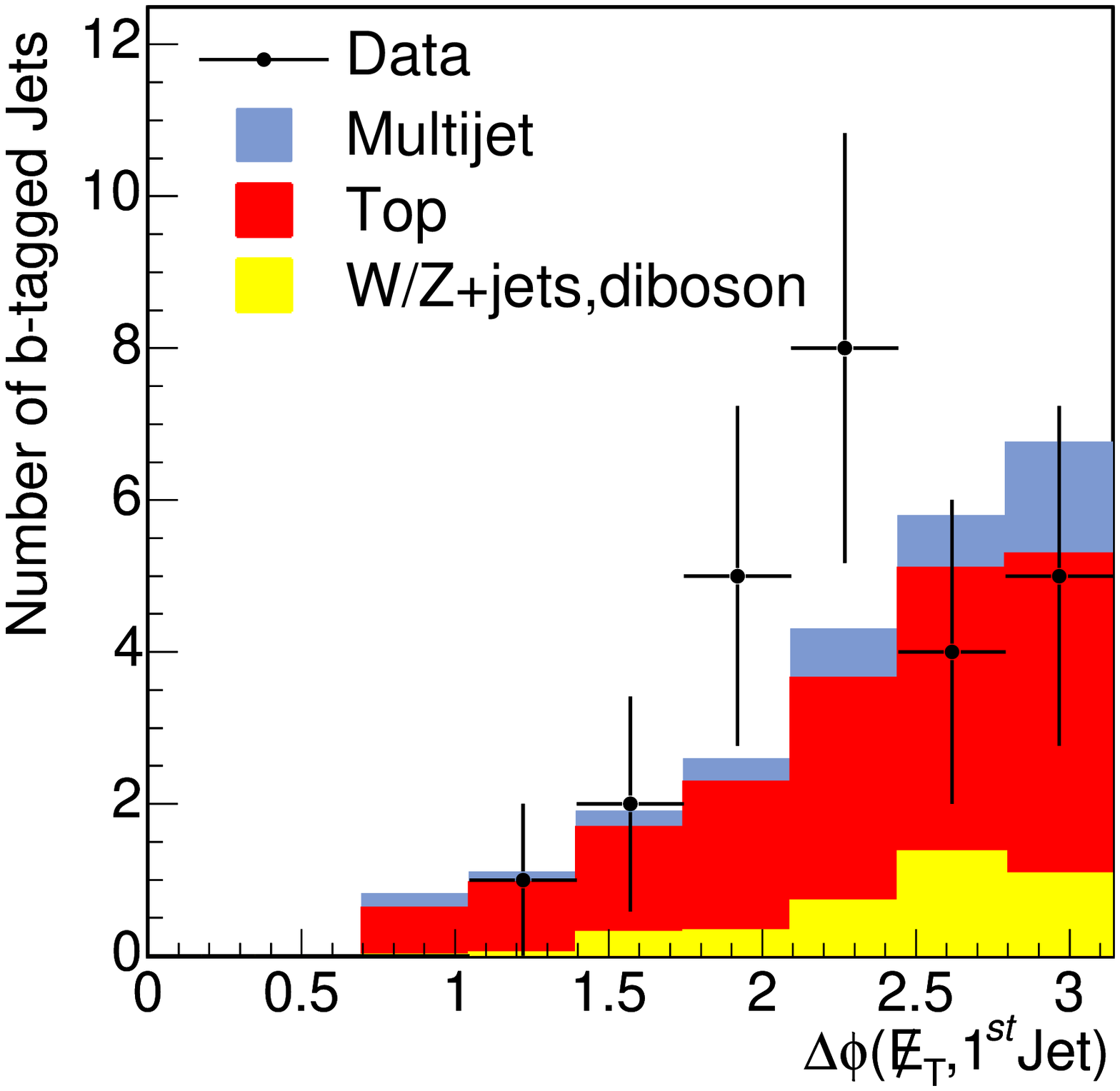}
\caption{ Distributions of $\Delta\phi(\met,\mbox{$1^{\mbox{\rm
st}}$ jet})$, the azimuthal angle between the $\met$ and the
highest $E_T$ jet in the event, if it is $b$-tagged in control
region C0 (left), C1 (middle), and C2 (right). } \label{dphi}
\end{figure*}

The data used for this search are selected by a three-level
trigger system. The Level 1 trigger requires $\met \ge 25\gev$,
where the $\met$ is estimated using trigger towers with more than
$1$~GeV of total $E_T$. The Level 2 trigger selects events with
two jet clusters with $E_T>10$~GeV. At Level $3$, the $\met$ is
required to be larger than $35$~GeV. After selections
~\protect\cite{Rott:2004nz} to remove accelerator-produced and
detector-related background, and cosmic ray events, 418,304 events
remain.  The position of the primary vertex along the beam axis,
$z_0$, is found by a weighted fit of each track's $z$ position at
closest approach to the beam axis, and must satisfy $|z_0|<60$~cm.
If there are multiple vertices in the event, we use the vertex
with the largest $\sum (p_T)$ of associated tracks. The events are
required to have $\met>35$~GeV, and at least three jets with
$|\eta|<2$ and $E_T \ge 15$~GeV. Events where the $\met$ is
aligned with any of the three highest $E_T$ jets are rejected by
requiring $\Delta\phi(\met,\mbox{1-3jet})>\frac{2\pi}{9}$ where
$\Delta\phi$ is the azimuthal angle between the $\met$ and each of
the jets. The last selection criterion reduces backgrounds with
$\met$ resulting from jet mismeasurements or from semileptonic
$b$-decays. The long lifetime of B hadrons yields secondary
vertices which have a large positive decay distance
$L_{xy}$~\protect\cite{l2d} relative to the primary vertex. We
require the events to have at least one jet identified (inclusive
single $b$-tagged) as a $b$-jet by the CDF $b$-tagging
algorithm~\protect\cite{Acosta:2004hw}. After applying these
selections, $332$ events are observed in the data.
\begin{table*}[b,h,t]
\caption{Comparison of the total number of expected and observed
inclusive single $b$-tagged events in the control regions. Only
statistical uncertainties are quoted.}
    \label{tab:table1}
    \begin{ruledtabular}
    \begin{tabular}{lrrr}
  Control region          & C0            & C1 & C2   \\ \hline
        $\met$                          & $35-50\gev$       & $ 35-50\gev$ & $>50\gev$  \\
        Isolated high-$p_T$ lepton  & required      & vetoed & required \\ \hline
        $W$/$Z$+jets/diboson        & $   3.9 \pm 0.8  $    & $  10.9 \pm 1.2  $ & $   9.6 \pm 1.2 $ \\
        Top             & $  11.7 \pm 0.2  $    & $  8.2 \pm 0.1  $ & $  35.0 \pm 0.3 $ \\
        Multijet                    & $  19.1 \pm 4.1  $    & $ 128.9 \pm 17.3  $ & $   10.9 \pm 4.5 $ \\ \hline \hline
        Total predicted             & $  34.7 \pm 4.2  $    & $ 148.0 \pm 17.3 $ & $   55.5 \pm 4.7 $ \\ \hline
        Observed                        & $  36   $     & $ 121   $ & $   63  $
\\
    \end{tabular}
    \end{ruledtabular}
\end{table*}

Dominant SM backgrounds are multijet, top, and
$W$/$Z$+jets/diboson (electroweak) production. The multijet
background is composed of a heavy flavor (HF) component and light
jets misidentified by the $b$-tagging algorithm as HF jets. The
latter are estimated from data using secondary vertices with
negative $L_{xy}$. The PYTHIA event
generator~\protect\cite{Sjostrand:2000wi} is used to estimate the
HF and top background. The top production is normalized to
theoretical predictions~\protect\cite{Cacciari:2003fi} while the
HF contribution is scaled to the data~\protect\cite{Rott:2004nz}.
The electroweak background is generated in
ALPGEN~\protect\cite{alpgen}, with the cross section at
next-to-leading order (NLO) calculated by the
MCFM~\protect\cite{mcfm} program, and the partons are showered in
the HERWIG~\protect\cite{Corcella:2000bw} program. The
CTEQ5L~\protect\cite{CTEQ5} parton distribution functions are used
in the generation of all background samples. Events are passed
through the standard simulation~\protect\cite{GEANT} of the CDF II
detector and weighted by the probability that they would pass the
trigger as determined in independent data samples.

To avoid potential bias when searching for new physics, we tested
the SM background in control regions that are defined {\it a
priori}. The data were divided into four regions having different
$\met$ and high-$p_T$ isolated lepton multiplicity. The signal
region contained events with large $\met$ and no high-$p_T$
isolated leptons. The three control regions used to check the SM
prediction, denoted as C0, C1, and C2, are defined in
Table~\ref{tab:table1}. C0 and C1 are sensitive to the predictions
of the multijet background. Control region C2 with high-$p_T$
isolated leptons and $\met>50$ GeV is used to check the top and
electroweak backgrounds. Predicted total number of events
and distributions of kinematic variables such
as the jet $E_T$, the track multiplicity and the $\met$ have been
studied and found to be in agreement with
observations~\protect\cite{Rott:2004nz} in the control regions. As
an example, Fig.~\ref{dphi} shows
$\Delta\phi(\met,\mbox{$1^{\mbox{\rm st}}$ jet})$, the azimuthal
angle between the $\met$ and the highest $E_T$ jet in the event
($1^{\mbox{\rm st}}$jet) if this jet is $b$-tagged. The
distributions show good agreement between data and prediction.
Table~\ref{tab:table1} summarizes the background contributions to
the number of expected inclusive single $b$-tagged events and the
observed events in the control regions.

\begin{figure}[t,h]
\includegraphics*[width=6.6cm]{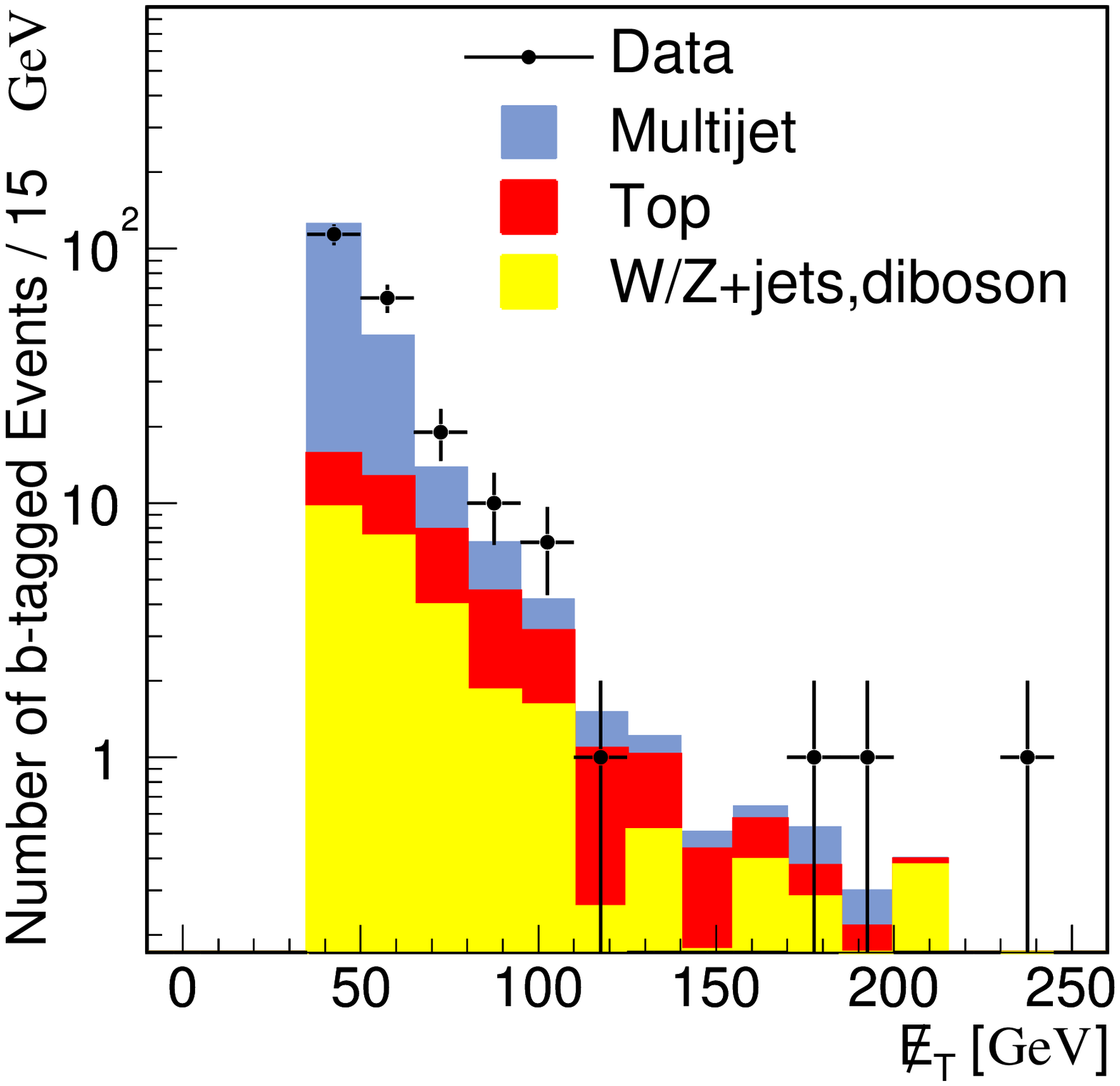}
\includegraphics*[width=6.6cm]{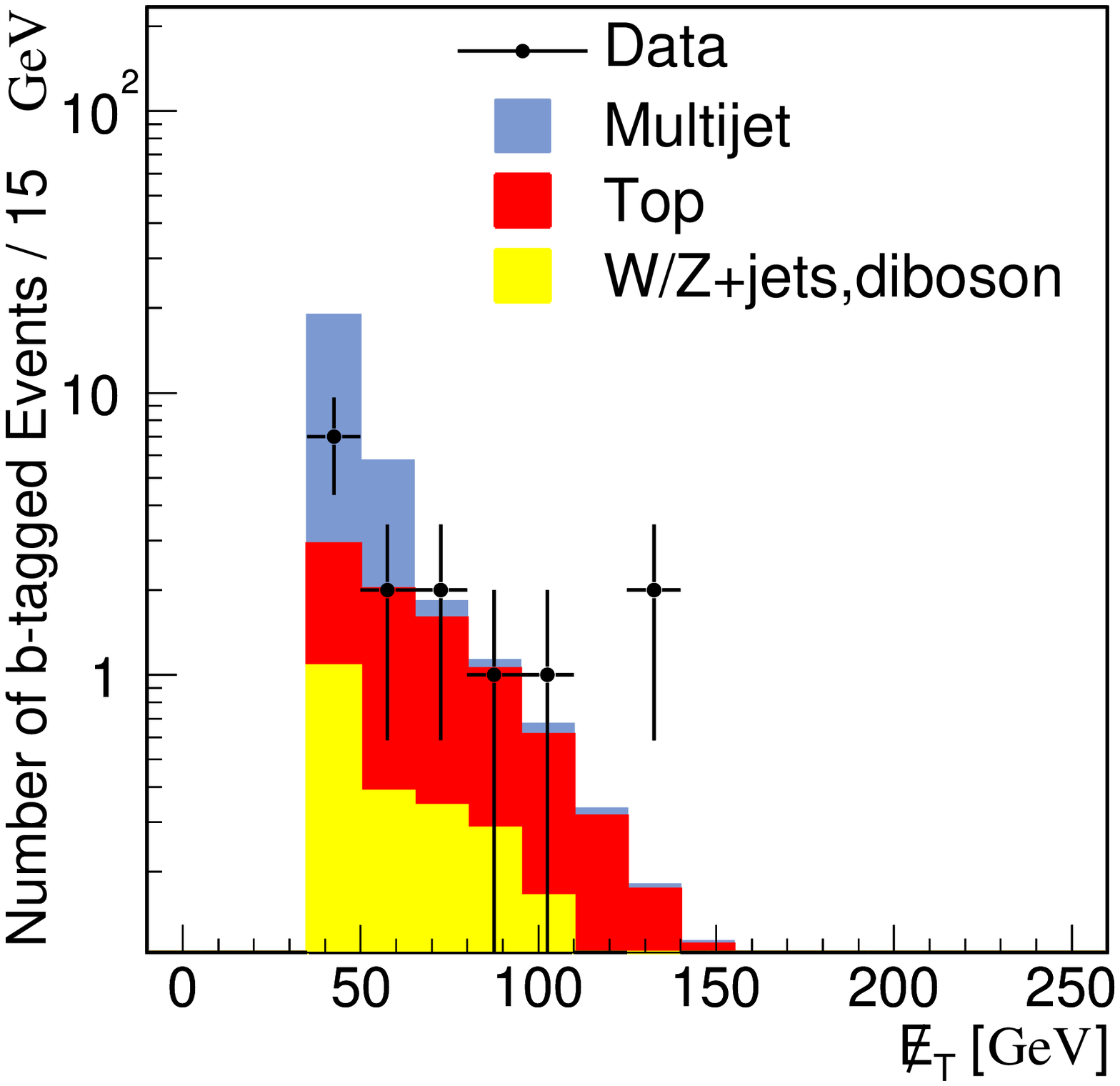}
\caption{$\met$ spectrum after vetoing events with high-$p_T$
isolated leptons, for single $b$-tagged events (top) and double
$b$-tagged events (bottom). } \label{fig:spectrums}
\end{figure}

We optimize the sensitivity to sbottom production from gluino
decays by maximizing the ratio of $N_{SUSY}/\sqrt{N_{SM}}$ where
$N_{SUSY}$ and $N_{SM}$ are the sbottom and SM background events
expected in the signal region~\protect\cite{Rott:2004nz},
respectively. The signal predictions are obtained by computing the
acceptance using the ISAJET~\protect\cite{isajet} event generator
normalized to the NLO production cross section determined with
PROSPINO~\protect\cite{Beenakker:1996ch} and the
CTEQ5M~\protect\cite{CTEQ5} parton distribution functions.
Figure~\ref{fig:spectrums} shows the $\met$ spectrum for events
with a single $b$-tagged jet (exclusive single $b$-tagged)
and at least two b-tagged jets in the event (inclusive double $b$-tagged)
when high-$p_T$ isolated leptons are vetoed.
The best sensitivity is achieved by requiring
inclusive double $b$-tagged events with $\met>80\gev$.
The double $b$-tag requirement is very effective at
suppressing the background.
Exclusive single tagged events were used to cross-check the results.
Since the efficiency of tagging a $b$-jet is about $30\%$~\protect\cite{Acosta:2004hw}, and there
are four $b$-jets in the signal final state, the signal acceptance
for single and double tag events are similar and
vary between $5\%$ and $12\%$ as a function of the sbottom and the gluino masses.

The systematic uncertainty on the signal and the
background predictions is studied in
detail~\protect\cite{Rott:2004nz}. Correlated and uncorrelated
uncertainties are evaluated separately and combined in quadrature.
Dominant uncertainties that affect both the background prediction
and the signal are the jet energy scale
($25\%$ for multijet and electroweak, $14.5\%$ for top,
$10\%$ for the signal), $b$-tagging efficiency ($12\%$),
and the luminosity ($6\%$).
Uncorrelated systematic uncertainties on the
background predictions are dominated by the heavy flavor MC scale
factor ($20\%$), the misidentified $b$-tag rate ($16\%$ for light
flavor multijets),  the top cross section ($15\%$), and the
electroweak cross section ($11.5\%$). The total systematic
uncertainties are $35\%$, $31\%$, and $25\%$ for the multijet,
electroweak and top background predictions, respectively and
$18\%$ on the signal.

The signal region is analyzed after all the background predictions
and selection cuts are determined. Requiring at least two
$b$-tagged jets we observed four events, where $2.6\pm 0.7$ are
expected, as summarized in Table~\ref{tab:table2}. We observe 21
exclusive single $b$-tagged events, which is in agreement with SM
background expectations of $16.3\pm 3.6$ events.

\begin{table}[t,h]
\caption{\label{tab:table2} Number of expected and observed events
in the signal region. Correlated and uncorrelated uncertainties
in the total predicted background were treated separately and combined
as described in~\protect\cite{Rott:2004nz}.}
\begin{ruledtabular}
\begin{tabular}{lcr}
Process     & Excl. Single $b$-Tag    & Incl. Double $b$-Tag\\
\hline
$W$/$Z$+jets/diboson & $    5.6 \pm 1.8     $ & $  0.6 \pm 0.3        $ \\
Top              & $    6.1 \pm 1.5             $ & $  1.9 \pm 0.5        $ \\
Multijet     & $    4.6 \pm 1.7         $ & $  0.2 \pm 0.1 $ \\
\hline
Total predicted  & $    16.3 \pm 3.6            $ & $  2.6 \pm 0.7        $ \\
\hline \hline
Observed         & $    21                      $ & $  4 $ \\
\end{tabular}
\end{ruledtabular}
\end{table}

\begin{figure}[t,h]
\includegraphics*[width=7.5cm]{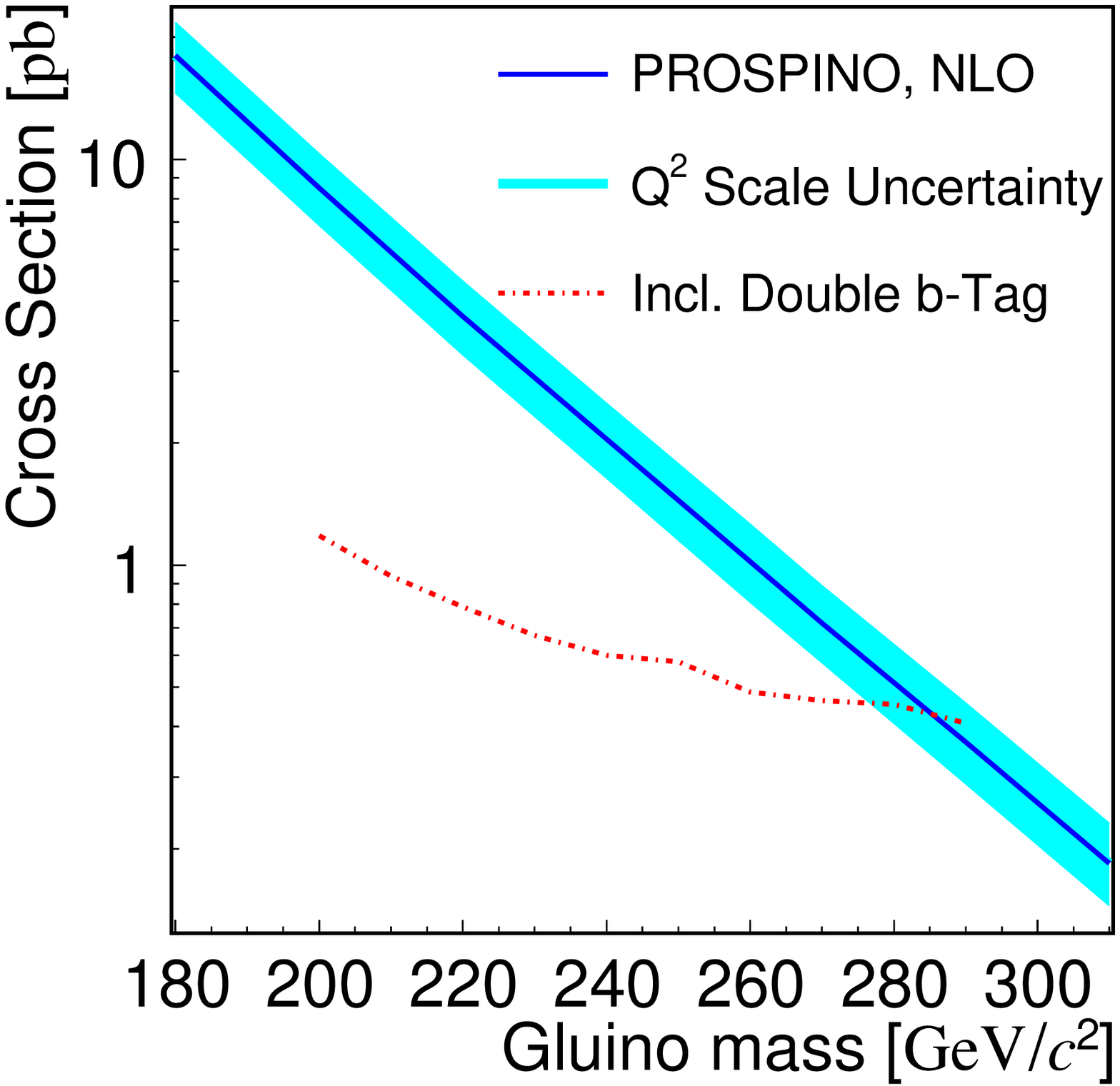}
\includegraphics*[width=7.5cm]{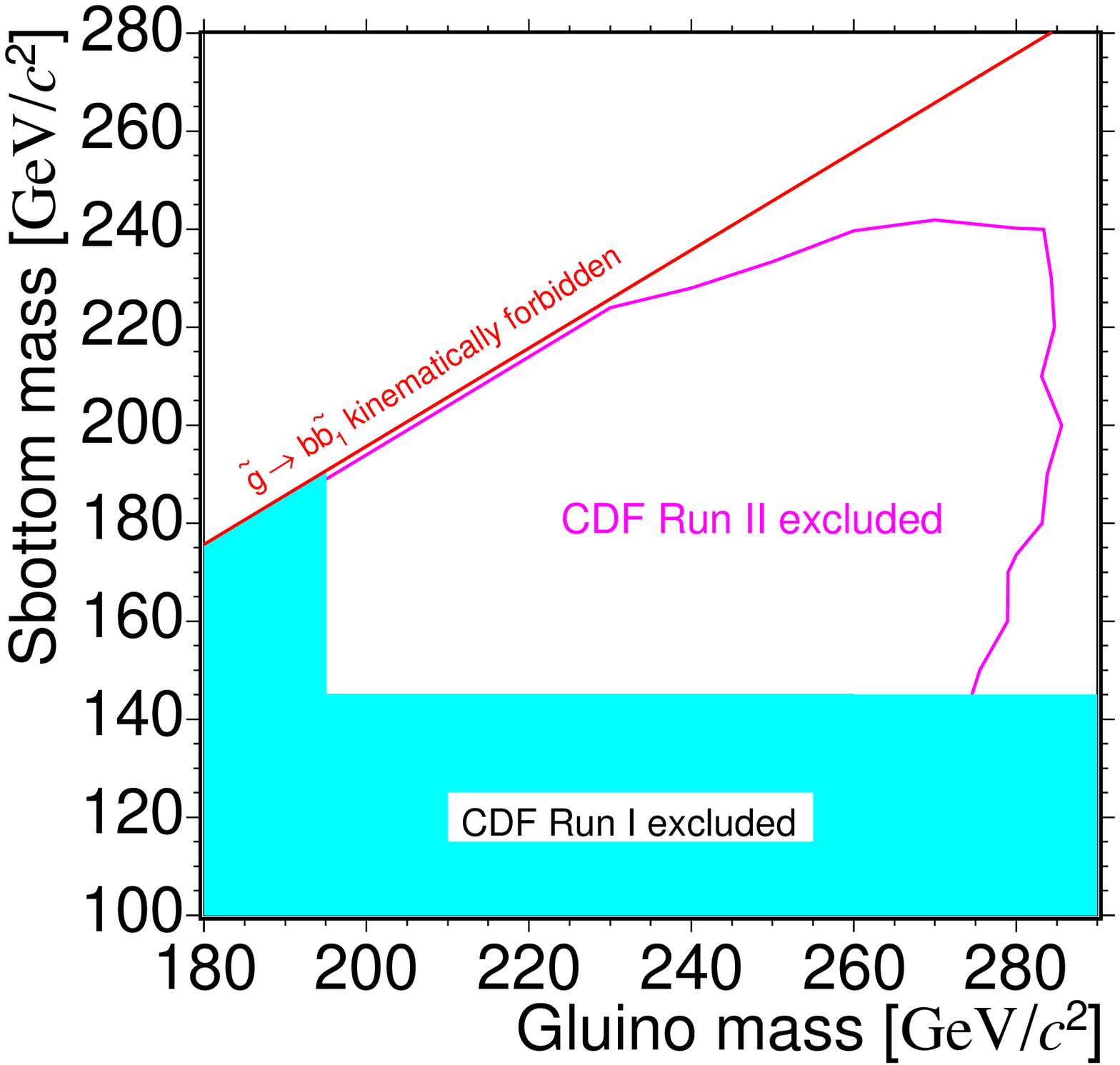}
\caption
{
           The 95\% C.L. cross section upper limit as a function of the gluino mass
           for a $60\gevm$ mass difference between the gluino and sbottom (top).
           The 95\% C.L. limit curve in the
           $m(\tilde{g})$-$m(\tilde{b_1})$ plane for inclusive double $b$-tagged events (bottom). }
\label{fig:exclusion}
\end{figure}

Since no evidence for gluino pair production with
$\tilde{g} \rightarrow b\tilde{b}_1$ is found, a cross section upper limit at
$95\%$ confidence level (C.L.) is computed and an exclusion limit
set using the Bayesian likelihood method~\protect\cite{limit} with
flat prior probability for the signal cross section and Gaussian
priors for the uncertainties on acceptance and backgrounds. An
example cross section upper limit is shown in
Fig.~\ref{fig:exclusion}. The contour exclusion limit shown in
Fig.~\ref{fig:exclusion} is computed using the inclusive
double $b$-tag analysis.
The $\tilde{\chi}_1^0$ mass has a minor
impact~\protect\cite{Rott:2004nz} on this limit, as long as the $\tilde{\chi}_1^0$ is sufficiently
lighter than the $\tilde{b}_1$, so that its decay produces a taggable $b$-jet.

In conclusion, we have searched for sbottom quarks from gluino
decays in $156$~pb$^{-1}$ of CDF Run II data. We observe four
inclusive double $b$-tagged candidate events, which is in
agreement with SM background expectations of $2.6\pm 0.7$ events.
No evidence for sbottom quarks from gluino decays is observed and
we exclude a significant region in the gluino and sbottom
mass plane at $95\%$ confidence level.
Gluino masses below $270\gevm$ for $\Delta m = m(\tilde{g}) - m(\tilde{b_1}) > 6\gevm$
and  $m(\tilde{b_1}) < 220\gevm$ are excluded.

We thank the Fermilab staff and the technical staffs of the
participating institutions for their vital contributions.
This work was supported by the U.S. Department of Energy and National
Science Foundation; the Italian Istituto Nazionale di Fisica
Nucleare; the Ministry of Education, Culture, Sports, Science and
Technology of Japan; the Natural Sciences and Engineering Research
Council of Canada; the National Science Council of the Republic of
China; the Swiss National Science Foundation; the A.P. Sloan
Foundation; the Bundesministerium f\"ur Bildung und Forschung,
Germany; the Korean Science and Engineering Foundation and the
Korean Research Foundation; the Particle Physics and Astronomy
Research Council and the Royal Society, UK; the Russian Foundation
for Basic Research; the Comisi\'on Interministerial de Ciencia y
Tecnolog\'{\i}a, Spain; in part by the European Community's Human
Potential Programme under contract HPRN-CT-2002-00292; and the
Academy of Finland.

\bibliography{apssamp}

\end{document}

%% file: July_2005_Authors_and_Siegrist.tex
\affiliation{Institute of Physics, Academia Sinica, Taipei, Taiwan 11529, Republic of China} 
\affiliation{Argonne National Laboratory, Argonne, Illinois 60439} 
\affiliation{Institut de Fisica d'Altes Energies, Universitat Autonoma de Barcelona, E-08193, Bellaterra (Barcelona), Spain} 
\affiliation{Baylor University, Waco, Texas  76798} 
\affiliation{Istituto Nazionale di Fisica Nucleare, University of Bologna, I-40127 Bologna, Italy} 
\affiliation{Brandeis University, Waltham, Massachusetts 02254} 
\affiliation{University of California, Davis, Davis, California  95616} 
\affiliation{University of California, Los Angeles, Los Angeles, California  90024} 
\affiliation{University of California, San Diego, La Jolla, California  92093} 
\affiliation{University of California, Santa Barbara, Santa Barbara, California 93106} 
\affiliation{Instituto de Fisica de Cantabria, CSIC-University of Cantabria, 39005 Santander, Spain} 
\affiliation{Carnegie Mellon University, Pittsburgh, PA  15213} 
\affiliation{Enrico Fermi Institute, University of Chicago, Chicago, Illinois 60637} 
\affiliation{Joint Institute for Nuclear Research, RU-141980 Dubna, Russia} 
\affiliation{Duke University, Durham, North Carolina  27708} 
\affiliation{Fermi National Accelerator Laboratory, Batavia, Illinois 60510} 
\affiliation{University of Florida, Gainesville, Florida  32611} 
\affiliation{Laboratori Nazionali di Frascati, Istituto Nazionale di Fisica Nucleare, I-00044 Frascati, Italy} 
\affiliation{University of Geneva, CH-1211 Geneva 4, Switzerland} 
\affiliation{Glasgow University, Glasgow G12 8QQ, United Kingdom} 
\affiliation{Harvard University, Cambridge, Massachusetts 02138} 
\affiliation{Division of High Energy Physics, Department of Physics, University of Helsinki and Helsinki Institute of Physics, FIN-00014, Helsinki, Finland} 
\affiliation{University of Illinois, Urbana, Illinois 61801} 
\affiliation{The Johns Hopkins University, Baltimore, Maryland 21218} 
\affiliation{Institut f\"{u}r Experimentelle Kernphysik, Universit\"{a}t Karlsruhe, 76128 Karlsruhe, Germany} 
\affiliation{High Energy Accelerator Research Organization (KEK), Tsukuba, Ibaraki 305, Japan} 
\affiliation{Center for High Energy Physics: Kyungpook National University, Taegu 702-701; Seoul National University, Seoul 151-742; and SungKyunKwan University, Suwon 440-746; Korea} 
\affiliation{Ernest Orlando Lawrence Berkeley National Laboratory, Berkeley, California 94720} 
\affiliation{University of Liverpool, Liverpool L69 7ZE, United Kingdom} 
\affiliation{University College London, London WC1E 6BT, United Kingdom} 
\affiliation{Massachusetts Institute of Technology, Cambridge, Massachusetts  02139} 
\affiliation{Institute of Particle Physics: McGill University, Montr\'{e}al, Canada H3A~2T8; and University of Toronto, Toronto, Canada M5S~1A7} 
\affiliation{University of Michigan, Ann Arbor, Michigan 48109} 
\affiliation{Michigan State University, East Lansing, Michigan  48824} 
\affiliation{Institution for Theoretical and Experimental Physics, ITEP, Moscow 117259, Russia} 
\affiliation{University of New Mexico, Albuquerque, New Mexico 87131} 
\affiliation{Northwestern University, Evanston, Illinois  60208} 
\affiliation{The Ohio State University, Columbus, Ohio  43210} 
\affiliation{Okayama University, Okayama 700-8530, Japan} 
\affiliation{Osaka City University, Osaka 588, Japan} 
\affiliation{University of Oxford, Oxford OX1 3RH, United Kingdom} 
\affiliation{University of Padova, Istituto Nazionale di Fisica Nucleare, Sezione di Padova-Trento, I-35131 Padova, Italy} 
\affiliation{LPNHE-Universite de Paris 6/IN2P3-CNRS} 
\affiliation{University of Pennsylvania, Philadelphia, Pennsylvania 19104} 
\affiliation{Istituto Nazionale di Fisica Nucleare Pisa, Universities of Pisa, Siena and Scuola Normale Superiore, I-56127 Pisa, Italy} 
\affiliation{University of Pittsburgh, Pittsburgh, Pennsylvania 15260} 
\affiliation{Purdue University, West Lafayette, Indiana 47907} 
\affiliation{University of Rochester, Rochester, New York 14627} 
\affiliation{The Rockefeller University, New York, New York 10021} 
\affiliation{Istituto Nazionale di Fisica Nucleare, Sezione di Roma 1, University of Rome ``La Sapienza," I-00185 Roma, Italy} 
\affiliation{Rutgers University, Piscataway, New Jersey 08855} 
\affiliation{Texas A\&M University, College Station, Texas 77843} 
\affiliation{Istituto Nazionale di Fisica Nucleare, University of Trieste/\ Udine, Italy} 
\affiliation{University of Tsukuba, Tsukuba, Ibaraki 305, Japan} 
\affiliation{Tufts University, Medford, Massachusetts 02155} 
\affiliation{Waseda University, Tokyo 169, Japan} 
\affiliation{Wayne State University, Detroit, Michigan  48201} 
\affiliation{University of Wisconsin, Madison, Wisconsin 53706} 
\affiliation{Yale University, New Haven, Connecticut 06520} 
\author{A.~Abulencia}
\affiliation{University of Illinois, Urbana, Illinois 61801}
\author{D.~Acosta}
\affiliation{University of Florida, Gainesville, Florida  32611}
\author{J.~Adelman}
\affiliation{Enrico Fermi Institute, University of Chicago, Chicago, Illinois 60637}
\author{T.~Affolder}
\affiliation{University of California, Santa Barbara, Santa Barbara, California 93106}
\author{T.~Akimoto}
\affiliation{University of Tsukuba, Tsukuba, Ibaraki 305, Japan}
\author{M.G.~Albrow}
\affiliation{Fermi National Accelerator Laboratory, Batavia, Illinois 60510}
\author{D.~Ambrose}
\affiliation{Fermi National Accelerator Laboratory, Batavia, Illinois 60510}
\author{S.~Amerio}
\affiliation{University of Padova, Istituto Nazionale di Fisica Nucleare, Sezione di Padova-Trento, I-35131 Padova, Italy}
\author{D.~Amidei}
\affiliation{University of Michigan, Ann Arbor, Michigan 48109}
\author{A.~Anastassov}
\affiliation{Rutgers University, Piscataway, New Jersey 08855}
% ...
\collaboration{CDF Collaboration}
\noaffiliation

%% file: Gluino_final_4.bbl
\begin{thebibliography}{99}
\bibitem{l2d}
$L_{xy}$ is the $r-\phi$ projection onto the jet axis of the
vector pointing from the primary vertex to the secondary vertex.
\end{thebibliography}
